\documentclass[twocolumn]{revtex4-1}
\usepackage{graphicx}
\usepackage{amsmath}
\usepackage{amssymb}
\usepackage{minibox}
\usepackage[usenames,dvipsnames,svgnames,table]{xcolor}
\usepackage{ctable}
\usepackage{microtype}
\usepackage[explicit]{titlesec}
\usepackage{lipsum}

\definecolor{lightgray}{gray}{0.93}
\titleformat{\subsection}
  {\normalfont \vspace{-0.5em}}{\thesection}{0em}{\normalsize\bfseries\sffamily #1 \vspace{-0.7em}}

\begin{document}

\title{Measuring absolute frequencies beyond the GPS limit via long-haul optical frequency dissemination }

\author{C.~Clivati$^{*\,1}$, G.~Cappellini$^{*\,2}$, L.~Livi$^{2}$, F.~Poggiali$^{2}$, M.~Siciliani~de~Cumis$^{1,5}$, M.~Mancini$^{3}$, G.~Pagano$^{3}$, M.~Frittelli$^{1}$, A.~Mura$^{1}$, G.~A.~Costanzo$^{4,1}$, F.~Levi$^{1}$, D.~Calonico$^{1}$, L.~Fallani$^{3,2}$, J.~Catani$^{5,2}$, M.~Inguscio$^{3,2,1}$}

\affiliation{
$^1$\minibox{INRIM Istituto Nazionale di Ricerca Metrologica, I-10135 Torino, Italy}\\
$^2$\minibox{LENS European Laboratory for Nonlinear Spectroscopy, I-50019 Sesto Fiorentino, Italy}\\
$^3$\minibox{Department of Physics and Astronomy, University of Florence, I-50019 Sesto Fiorentino, Italy}\\
$^4$\minibox{Politecnico di Torino - Dipartimento di Elettronica e Telecomunicazioni, I-10129 Torino, Italy}\\
$^5$\minibox{INO-CNR Istituto Nazionale di Ottica del CNR, Sezione di Sesto Fiorentino, I-50019 Sesto Fiorentino, Italy}\\
}

\begin{abstract}
{\bf \noindent
Global Positioning System (GPS) dissemination of frequency standards is ubiquitous at present, providing the most widespread time and frequency reference for the majority of industrial and research applications worldwide. On the other hand, the ultimate limits of the GPS presently curb further advances in high-precision, scientific and industrial applications relying on this dissemination scheme. Here, we demonstrate that these limits can be reliably overcome even in laboratories without a local atomic clock by replacing the GPS with a 642-km-long optical fiber link to a remote primary caesium frequency standard. Through this configuration we stably address the $^1$S$_0$---$^3$P$_0$ clock transition in an ultracold gas of $^{173}$Yb, with a precision that exceeds the possibilities of a GPS-based measurement, dismissing the need for a local clock infrastructure to perform high-precision tasks beyond GPS limit. We also report an improvement of two orders of magnitude in the accuracy on the transition frequency reported in literature.
}
\end{abstract}

\maketitle

\noindent
Frequency and time dissemination is of fundamental importance for the modern world, as it rules a wealth of applications, ranging from coordination of global transport to synchronization of high-precision tasks in both scientific and industrial sectors. A timely exploitation of the continuous advances in the realization of new frequency standards is inherently tied to the uncertainty with which the standard is delivered from metrological institutes to the end-users. Presently, the most widespread dissemination method is the GPS, allowing for traceability to the second in the International System (SI) of units with a typical fractional frequency uncertainty of $10^{-13}$ on a day average \cite{ref1}. This remarkable result can be obtained with a very simple equipment, consisting only of a receiver without any feedback from the user to the reference clock infrastructure. However, GPS dissemination severely degrades the accuracy of current frequency standards and even more that of the optical atomic clocks \cite{ref3}, which have already shown fractional accuracy capabilities of $10^{-18}$ in few hours of measurement \cite{ref4,ref5,ref6}. This is the main reason why the development and characterization of long-haul optical fiber links (OFLs), rather than microwave satellite links, recently featured a substantial thrust \cite{epl2015,ref7,ref8,ref9,ref10}. Indeed, in the last years OFLs have shown undoubted effectiveness in comparing remote atomic clocks beyond the ultimate limit of GPS-based comparisons, attaining precision levels not limited by the frequency transfer method \cite{ref11,ref12,ref13,ref14,lisdat15}, and are also envisioned as key elements for the development of a new kind of high-accuracy relativistic geodesy \cite{ref15, lisdat15}. Besides optical clock comparisons, short-range OFLs have recently been exploited in order to increase the spectral performances of a local laser and to allow for precision molecular spectroscopy \cite{ref16}.

\begin{figure*}[t!]
\begin{center}
\includegraphics[width=0.93\textwidth]{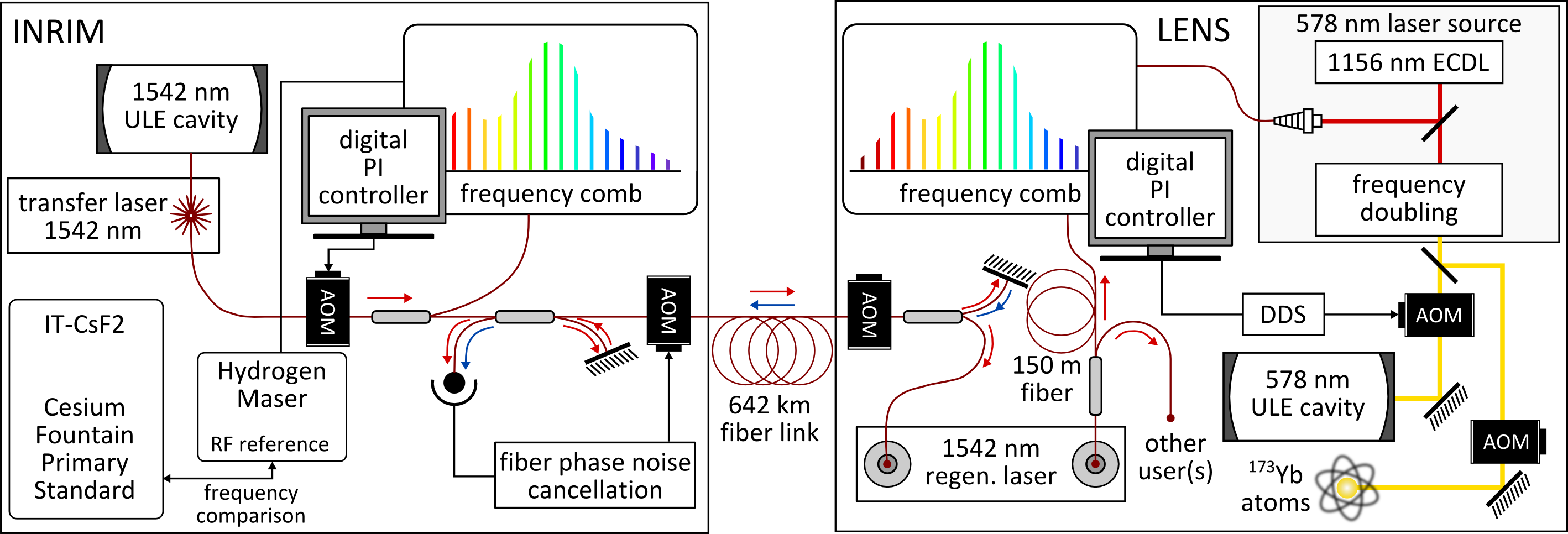}
\end{center}
\caption{{\bf Sketch of the experimental system.} Left panel: generation of the optical signal, frequency-referenced to the Cs primary standard at the metrological institute INRIM (Turin), for long-haul optical frequency dissemination. Right panel: end-user spectroscopy system at LENS (Florence), along with signal regeneration and frequency bridging with a local frequency comb. Note that other users can exploit the reference after regeneration due to a ``point-to-star'' dissemination configuration.} \label{fig1}
\end{figure*}

However, there is still no evidence about their successful utilization to disseminate a primary absolute frequency reference to remote laboratories lacking a local atomic clock, in such a way as to push both precision and accuracy of local measurements and applications beyond the GPS limit. A demonstration of such a capability would represent a breakthrough, as it would allow remote, non-metrological end-users to realize tasks requiring an ultra-high degree of accuracy, which are unattainable with GPS-disciplined standards. Focusing on possible scientific outcomes of this new approach, a paradigmatic example comes from experiments involving alkaline-earth-like (AEL) atoms. Very recently, these systems proved to lie amongst the most promising and rich platforms for the development of new quantum technologies. Coherent control on their electronic and nuclear degrees of freedom allows for the realization of quantum simulators of a wealth of fundamental effects which are hardly attainable (if not unfeasible) in their original physical context, with the investigation of orbital quantum magnetism \cite{gorshkov2010}, SU($N$) fermionic systems \cite{cazalilla2014, taie2012, pagano2014}, synthetic gauge fields \cite{mancini2015}, as well as the implementation of reliable quantum information schemes \cite{daley2008,daley2011}. Much of the interest sought in this class of atoms resides in the possibility to handle their orbital degree of freedom through the forbidden transition $^1$S$_0$---$^3$P$_0$ \cite{cappellini2014,scazza2014,zhang2014,pagano2015,hofer2015}, that is also exploited for the realization of optical lattice clocks with neutral atoms \cite{ref3}. However, most of the aforementioned experimental schemes rely on cyclical addressing of the clock transition (featuring linewidths of few Hz) for hours or even days, creating the need for manipulation lasers with narrow linewidth and exceptional long-term stability. Whilst the first requirement is easily met through high-finesse cavity stabilization, long-term stability can only be achieved through referencing to atomic clocks, whose availability is primarily restricted to metrological institutes.

Here, we demonstrate that the excellent long-term stability of an ultrastable near-infrared laser, stabilized to an atomic clock, can be transferred from a metrological institute to a remote, non-metrological research laboratory through a 642-km-long optical fiber link. This allows for a long-term, SI-traceable spectroscopy of the $^1$S$_0$---$^3$P$_0$ transition in a quantum degenerate AEL gas of $^{173}$Yb. The absolute measurement of the $^{173}$Yb $^1$S$_0$---$^3$P$_0$ transition frequency is $518\,294\,576\,845\,268\,(10)$ Hz on typical timescales of few hours, and improves the previously known value by two orders of magnitude \cite{ref28}. Noticeably, this degree of accuracy exceeds by 20 times the peformances attainable with a GPS-based frequency transfer on the same timescale \cite{ref1}. Our results pave the way for a new generation of optically-referenced high-precision setups beyond the GPS limit showing no need for a local optical clock infrastructure.

\subsection*{Optical frequency link: setup and performances}

\noindent
The complete system (Cs fountain clock + optical frequency reference + optical link + remote spectroscopic setup) is shown in Fig. 1.  An optical frequency reference is generated in a national metrological institute (INRIM, Turin, Italy) by frequency-stabilizing a 1542-nm fiber laser to a  Ultra Low Expansion glass (ULE) cavity \cite{ref31}. On the long-term, the laser is phase-locked to a hydrogen maser (HM) referenced to a Cs fountain primary frequency standard with $2 \times 10^{-16}$ relative frequency accuracy \cite{ref32}. This ensures excellent long-term stability and SI-traceability (see Methods). The optical reference is disseminated from INRIM to LENS via a phase-stabilized 642-km-long fiber link characterized at the $5 \times 10^{-19}$ level of uncertainty \cite{ref8}. The occurrence of occasional phase-slips is continuously monitored to keep undesired frequency offsets $<10^{-16}$ on the delivered frequency (see Methods). At LENS, the incoming optical reference is regenerated and fiber-delivered to the end-user lab (see Methods). Multiple amplified output slots are available, realizing a ``point-to-star'' configuration, where other end-users could also connect to the regeneration stage, either with free-running or stabilized links upon specific metrological needs. In the end-user lab, a quantum-dot laser at 1156 nm is frequency stabilized to a high-finesse ULE cavity \cite{ref29,ref30}. Although this cavity is enclosed in a thermally-insulating box, the stabilized laser shows stochastic frequency fluctuations mainly arising from the non-perfect thermal stabilization of the ULE cavity, in addition to a linear (aging) drift of about 5 kHz/day \cite{ref30}. This instability is compensated by frequency-locking the 1156-nm laser to the disseminated radiation at 1542 nm, using an optical frequency comb as a bridge between the two spectral regions. The complete metrological chain provides the end-user with a 1156-nm laser frequency traceable to the SI second, realized by the Cs fountain; the 1156-nm laser keeps its own high spectral purity on the short term, while the long-term stability is granted by the fiber-disseminated signal on measurement times of tens of seconds. The 1156-nm laser radiation is then frequency-doubled to 578 nm and used for spectroscopy of the $^1$S$_0$--$^3$P$_0$ clock transition of an ultracold sample of $^{173}$Yb.

\begin{figure*}[t!]
\begin{center}
\includegraphics[width=0.93\textwidth]{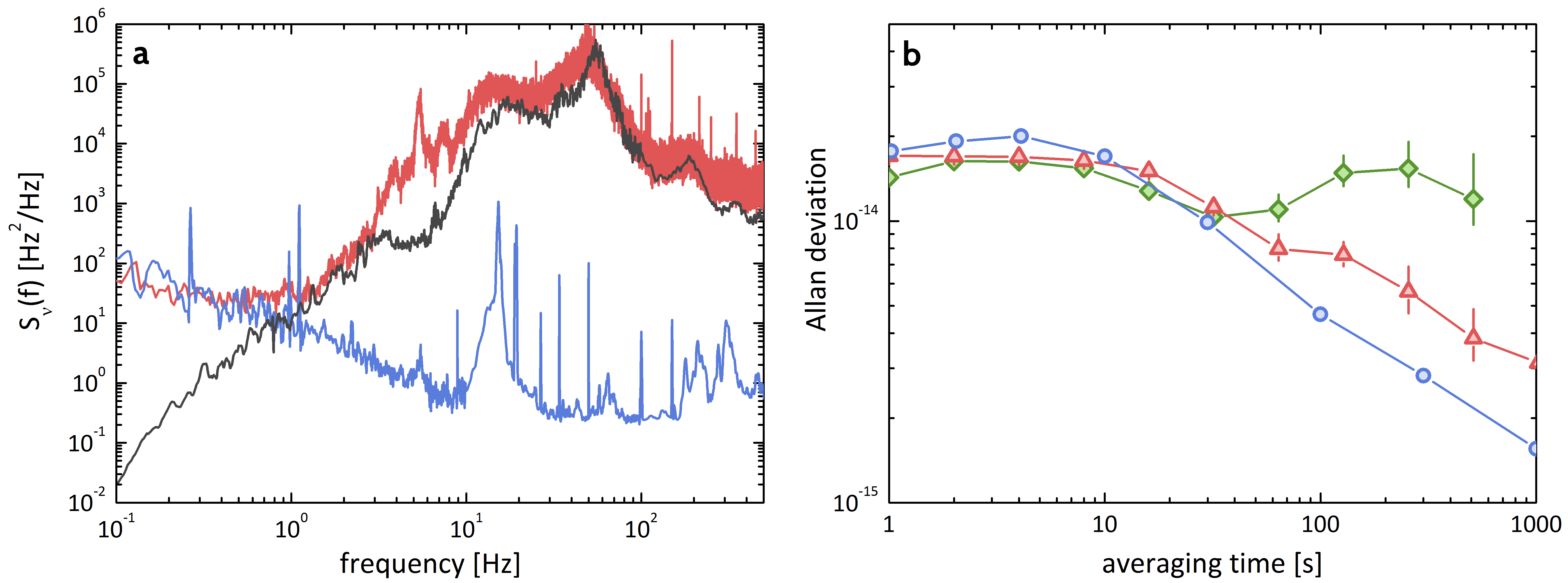}
\end{center}
\caption{{\bf Performances of the optical frequency link stabilization.} {\bf a.} Red line: frequency noise of the beatnote between the local laser at 1156 nm and the reference laser at 1542 nm; blue line: frequency noise of the HM-disciplined 1542-nm laser as measured at INRIM; black line: expected contribution of the optical link. All the spectra are referred to the 1156-nm spectral region. {\bf b.} Blue line (circles): fractional frequency instability (Allan deviation) of the HM-disciplined 1542-nm laser as measured at INRIM; red line (triangles): instability of the beatnote between the fiber-disseminated signal at 1542 nm and the local 1156-nm laser, locked to it; green line (diamonds): instability of the beatnote between the 1542-nm laser and the local 1156-nm laser when a linear drift of 0.1 Hz/s is removed off-line. All measurements are taken with 0.5 Hz measurement bandwidth. 
} \label{fig2}
\end{figure*}

Fig. 2{\bf a} shows the frequency noise spectrum of the employed optical sources referred to the 1156 nm spectral region. The blue line represents the frequency noise of the 1542-nm reference laser, disciplined to the HM, when compared to an independent, cavity-stabilized laser at INRIM. The red curve represents the frequency noise of the beatnote between the 1542-nm laser and the 1156-nm laser through the frequency-comb bridge. A significant noise increase can be noticed at Fourier frequencies higher than 1 Hz, which is due to the residual fluctuations of the phase-stabilized link. The black curve represents the link contribution alone, calculated from the round-trip fiber noise, considering that the compensation is limited by the photon round-trip time into the fiber ($\sim 6$ ms) \cite{ref33}. At Fourier frequencies higher than 0.1 Hz, the fiber-disseminated optical signal gives the main contribution to the noise and cannot be directly used for spectroscopy: this motivates the ULE-stabilization of the local 1156-nm laser. On the other hand, the long-term instability of the 1156-nm local laser emerges on long averaging times: in Fig. 2{\bf b}, the green curve shows the relative frequency instability (relative frequency Allan deviation) of the beatnote between the disseminated 1542-nm laser and the local 1156-nm laser, after a linear drift ($\sim 0.1$ Hz/s) is removed. At averaging times $>10$ s it is limited by the stochastic temperature variations of the ULE cavity to which the 1156-nm laser is locked. In this configuration, short-term instabilities of few parts in $10^{-14}$ are featured by the probe laser, still showing frequency drifts $\sim 100$ mHz/s. The red line shows the relative instability of the beatnote between the disseminated 1542-nm laser and the local 1156-nm laser, when the latter is locked to the former; this represents the electronic and system noise of the comb and locking algorithm at LENS and demonstrates that the long-term stability of the 1156-nm laser can be improved by one order of magnitude already at 1000 s, achieving the $10^{-15}$ level of instability. The blue line shows the instability of the 1542-nm laser disciplined to the HM, measured against an independent twin system at INRIM.

\subsection*{Beyond-GPS, optically-referenced spectroscopy}

\noindent
To demonstrate the potential of optical frequency dissemination in actual experiments, we performed long-term interrogation of the $^1$S$_0$--–$^3$P$_0$ clock transition (featuring a natural linewidth of few tens of mHz) in a quantum degenerate Fermi gas of $^{173}$Yb. The atomic sample is prepared by means of laser-cooling and evaporative-cooling techniques, reaching temperatures as low as 50 nK. In this regime, roughly $10^4$ (spin-polarized) atoms are loaded into a three-dimensional optical lattice, with one atom per lattice site, each being separated by 380 nm from the nearest neighbors. Due to the low temperature and fermionic nature of the particles, the Pauli exclusion principle inhibits multiply-occupied sites. The lattice depth is $\sim 30 E_\mathrm{rec}$, where $E_\mathrm{rec} = h \times 2.00$ kHz stands for the atomic recoil energy at the lattice wavelength of 759 nm and $h$ for the Planck's constant. The atoms are illuminated by a 100-ms-long pulse of 578-nm light, coming from the frequency-doubled 1156-nm probe laser. The $\pi$-polarized probe beam intensity is 10 $\mu$W/cm$^2$, adjusted in order to grant a reliable signal-to-noise ratio and leading to a line broadening of less than 50 Hz. The atoms are then released from the lattice, and the spectroscopic signal is extracted by observing the loss peak in the total atom number as a function of the laser frequency through resonant light absorption after a 25-ms-long time of flight.

\begin{figure*}[t!]
\begin{center}
\includegraphics[width=0.93\textwidth]{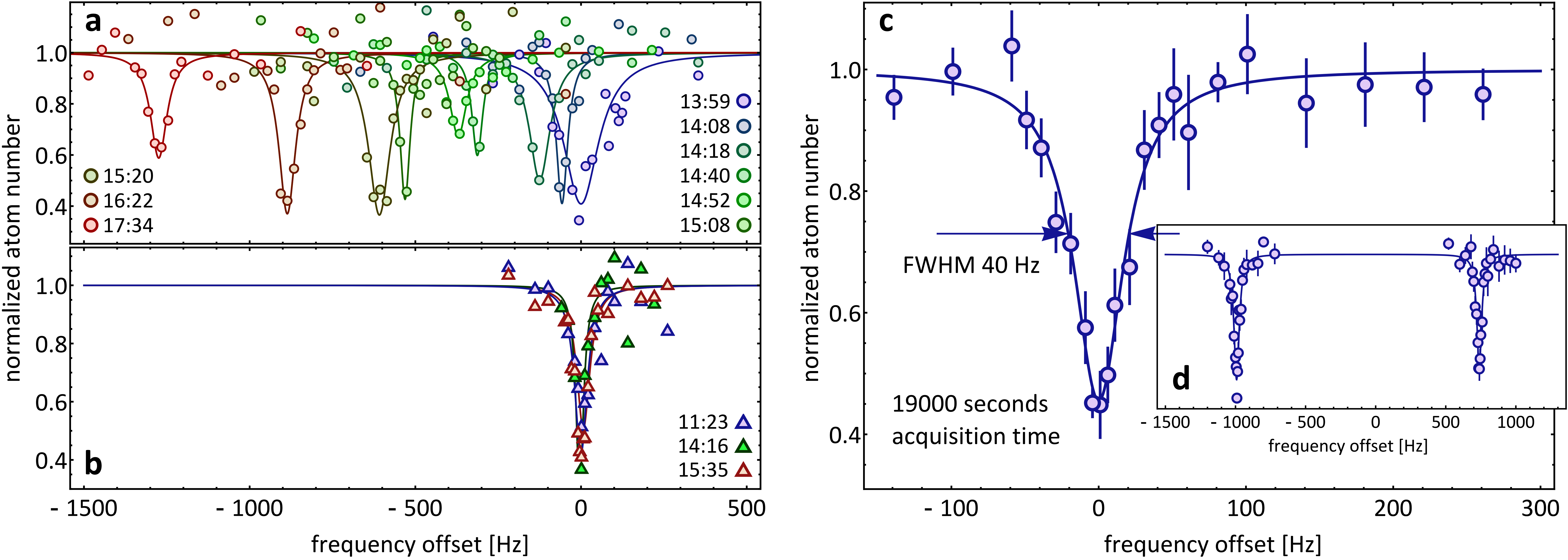}
\end{center}
\caption{{\bf Spectroscopy of the $^1$S$_0$---$^3$P$_0$ $^{173}$Yb clock transition.} {\bf a.} GPS-based measurement, with a series of 10- to 30-minutes-long scans taken at different starting times during the same day (see legend). {\bf b.} Same as in {\bf a}, but with the GPS reference replaced by the optical link. {\bf c, d.} Long-term addressing of the clock transition (5.2 hours), used for extracting its absolute frequency (see text). The error bars indicate the standard deviation of the mean over approximately 10 repeated measurements. The lines are fits with Lorentzian functions. The zero values for the horizontal scales are chosen arbitrarily.} \label{fig3}
\end{figure*}

Fig. 3{\bf a,b} shows typical short-term atom loss features retrieved by this method on a total timescale of several hours. Here, each scan takes typically 20 minutes, since the experimental setup is not optimized for fast-cycle, high-precision spectroscopy. In order to stress the benefits of the optical link referencing, we compare the case of (a) laser referenced to the GPS standard and (b) laser stabilized to the optical link. The difference is remarkable. In the GPS configuration case, drifts larger than 1 kHz are observed, being the linewidth/lineshape heavily influenced by the relative direction of scan and laser drifts on the observation timescale. The spectra show typical Lorentzian full-width-at-half-maximum (FWHM) linewidths below 100 Hz. On the other hand, in the optical link case no appreciable fluctuations/drifts arise. Here the residual laser fluctuations are well below the typically observed linewidth, not exceeding tens of Hz FWHM, limited by the short term stability of the local laser \cite{ref30}, which is comparable to the power broadening. 

The remote long-term stabilization of the interrogation laser is a keynote result for the vast majority of experimental setups that require a high degree of stability: the local sample can be employed as a ``tool'', rather than as a frequency reference (as happening, e.g, in optical clocks), and in turn the local laser can be used as a coherent manipulation device for the sample itself, rather than as a probe. Specifically, for experiments involving AEL atoms, this implies the possibility of building new quantum simulators involving the manipulation of the electronic degree of freedom, or realizing implementation of quantum computing protocols in a steady, reliable way. As a matter of fact, in a typical experimental setup as the one at LENS, which is designed for those tasks (and not as an optical atomic clock), a direct lock of the local laser frequency on the clock atomic transition is not a viable choice. Indeed, the experimental cycle (devoted to the manipulation of a quantum degenerate gas) can last from several seconds up to few minutes. If the local probe laser were locked on the atomic transition, this rather long cycle would cause a large deterioration of the laser frequency stability because of the Dick effect \cite{ref3}. 

\subsection*{Absolute frequency measurement of the $^{173}$Yb clock transition}

\noindent
By exploiting the optical dissemination, we have measured the absolute frequency of the $^1$S$_0$---$^3$P$_0$ clock transition with respect to the remote Cs fountain. We collected six independent measurements spread over a period of three months, each consisting of a complete scan of the transition, with typical linewidths of 50 Hz FWHM (see Fig. 3{\bf c}). The runs had different duration, ranging from 2000 s to 19000 s, for a total measurement time of 40000 s ($\sim 11$ hours). The measured frequency $\nu_0$ of the $^{173}$Yb $^1$S$_0$---$^3$P$_0$ transition is 518\;294\;576\;845\;268\;(10) Hz; this improves the uncertainty of a previous direct measurement \cite{ref28} by a factor 400, with an agreement between the two values. Consequently, combined with other previous measurements \cite{ref3}, we also improve the 
knowledge of the isotope shifts of the $^1$S$_0$---$^3$P$_0$ transition, whose values are: $\nu_0^{171}-\nu_0^{173}=1\,259\,745\,597(10)$ Hz and $\nu_0^{173}-\nu_0^{174}=551\,536\,050(10)$ Hz. Remarkably, this accuracy would not be reachable through a GPS-based measurement.

Table 1 reports the uncertainty budget, along with the systematic biases that have been corrected to obtain the unperturbed $^{173}$Yb clock transition frequency. We calculate the total uncertainty on the absolute frequency value as the quadratic sum of statistical and systematic contributions. The statistical uncertainty (Type A) is the weighted composition of four contributions: the Lorentzian fit, the Cs fountain instability and the beatnotes instability of the optical combs at INRIM and LENS. The total type A uncertainty for the reported measurement is 1.9 Hz ($4 \times 10^{-15}$ fractional frequency uncertainty), where we considered a Student statistic for the data and a confidence level of 90\% due to the limited number of measurements. 

The systematic uncertainty (Type B) considers three sources: the Cs fountain accuracy, the phase slips of the optical link and the physical effects perturbing the Yb clock transition. The contribution coming from the Cs fountain accuracy ($2 \times 10^{-16}$) was 0.1 Hz \cite{ref32}, whilst the contribution by the phase-slips on the fiber link was typically $<0.1$ Hz ($1.5 \times 10^{-16}$), measured by redundant beatnote tracking.

In order to provide an unbiased value for the absolute frequency of the $^{173}$Yb clock transition, it is necessary to evaluate the local systematics. A quantization magnetic field is applied in order to resolve the Zeeman structure of the line, arising from the $^{173}$Yb nuclear spin $I=5/2$. The transition frequency measurement is obtained by a Lorentzian fit of the two resonances corresponding to Zeeman components $m_I=\pm 5/2$, reported in Fig. 3{\bf d}, and then by averaging their observed values in order to wash out the first-order magnetic field contribution to the transition frequencies. The average first-order splitting of the two components is 1712(3) Hz, hence, according to the sensitivity coefficient of $2 I \times 113$ Hz/G \cite{derevianko}, the value of the bias magnetic field is $B=3.031(5)$ G. This value is used to correct for the quadratic Zeeman shift of the transition frequency. The quadratic sensitivity is experimentally measured to be $-0.064(2)$ Hz/G$^2$, leading to a bias of $-0.59(3)$ Hz in the measurements. 

\begin{table}[t!]
\caption{{\bf Uncertainty budget of the $^{173}$Yb $^1$S$_0$---$^3$P$_0$ absolute frequency, expressed in Hz at 578 nm.}}
\begin{center}
{\renewcommand{\arraystretch}{1.15}
\setlength{\fboxsep}{0pt}
\colorbox{lightgray}
{\begin{tabular*}{\columnwidth}{l @{\extracolsep{\fill}} cc}
\specialrule{.1em}{0pt}{0pt} 
{\bf Contribution} & {\bf Bias} & {\bf Uncertainty}\\
& (Hz) & (Hz)\\
\specialrule{.1em}{.1em}{.1em}
Lorentzian fit (*)& -- & $0.8$---$5$ \\
Cs fountain statistical (*)& -- & $0.9$---$2$ \\
Comb INRIM statistical (*)& -- & $0.4$---$1.2$ \\
Comb LENS statistical (*)& -- & $1$---$3$ \\
{\bf Total Type A (**)} & & {\bf 1.9} \\
\hline
% {\it Type B (dissemination):} & & \\
Cs fountain standard accuracy& -- & $0.1$ \\
Fiber link phase slips (***)& -- & $0.1$---$5$ \\
% {\it Type B (spectroscopy):} & & \\
Quadratic Zeeman & $-0.59$ & $0.03$ \\
Lattice Stark & -- & $8$ \\
Blackbody radiation & $-1.24$ & $0.05$ \\
Probe laser intensity & -- & $0.00015$ \\
Gravitational redshift & $2.277$ & $0.005$ \\
{\bf Total Type B (***)} & & {\bf 9} \\
\specialrule{.1em}{.1em}{.1em} 
{\bf Total (***)} & {\bf 0.5} & {\bf 10} \\
\specialrule{.1em}{0pt}{0pt} 
\end{tabular*}}}
\end{center}
\scriptsize
\begin{tabular*}{\columnwidth}{lp{0.9\columnwidth}}
(*) & Depending on the measurement run.\\
(**) & Weighted uncertainty of all measurements with Student 90\% confidence level.\\
(***) & Typically 0.1 Hz; in some measurements the phase-slips uncertainty was 5 Hz for technical problems; in the total type B we considered the worst-case scenario.
\end{tabular*}
\end{table}

In order to limit the effects of the differential light shift induced on the $^1$S$_0$ and $^3$P$_0$ levels by the trapping light potential, the optical lattice has been set at the Stark-shift-free ``magic'' wavelength around 759 nm \cite{ref3}. The value of the magic frequency has been measured by using the complete metrological chain as described in the previous section (see Methods). The experimental sensitivity of the atomic resonance frequency on the lattice detuning from the magic frequency is $-1.53(2)$ Hz/GHz for our lattice depth. During the absolute frequency measurement, the lattice frequency was monitored to keep the uncertainty contribution $<8$ Hz, resulting in the leading component of the uncertainty budget.

The blackbody radiation shift is accounted for by using the Yb sensitivity reported in \cite{ref35}. The atoms are probed in a glass science chamber (not facing the hot atomic source), that is at thermal equilibrium with room temperature. The resulting frequency bias is $-1.24(5)$ Hz for a room temperature of $298(3)$ K. 

The probe light induces a negligible light shift on the transition: according to the sensitivity of 15 mHz/mW\,cm$^{-2}$ \cite{poli08} and to the laser intensity of 10 $\mu$W/cm$^2$, the light shift is 0.15 mHz, kept as uncertainty contribution. 

Collisional shifts are also negligible, since the spin-polarized Fermi gas fills the 3D optical lattice with at most one atom per site.

The SI second is defined at the geoid gravity potential, hence the frequency reference disseminated from INRIM is corrected for the gravitational redshift \cite{ref36}. The orthometric height of the LENS laboratory on the geoid was measured in a dedicated geodetic campaign \cite{ref37}, and the resulting gravitational redshift is $4.81(1)\times 10^{-15}$ or 2.277(5) Hz (see Methods for details).

\subsection*{Conclusions}

\noindent
In conclusion, we have demonstrated that long-haul fiber-based optical frequency dissemination is a reliable tool for remote end-users to perform high-precision procedures well beyond the ultimate capabilities of GPS dissemination. As a proof, ultrastable laser light at 1542 nm, traced to the SI second, is transferred from the Italian National Metrology Institute to the European Laboratory for Non Linear Spectroscopy, across a 642-km-long fiber link. This is used as an absolute frequency reference for a metrological measurement in a non-metrological experiment at LENS, enabling a stable, long-term interrogation of the  sample with metrological traceability and the absolute measurement of the $^{173}$Yb $^1$S$_0$---$^3$P$_0$ clock transition frequency with a combined uncertainty of 10 Hz, that also improves the previously published result \cite{ref28} by a factor 400. The inherent uncertainty of the dissemination channel is limited to 2 Hz, corresponding to a relative uncertainty of $4 \times 10^{-15}$. Noticeably, this beyond-GPS accuracy is achieved in timescales as short as few hours and can be reliably reproduced over time periods of several months in a non-metrological end-user laboratory.

This long-haul optical frequency dissemination features a potential outcome on a wealth of applications ranging from scientific research to industrial development and production processes. Besides the intrinsic interest of our work as a proof of real effectiveness of this dissemination technique, our results could be readily exploited for the investigation of many-body quantum physics in AEL atomic systems \cite{gorshkov2010,zhang2015}, and for the application of novel promising quantum information schemes \cite{daley2008} where the frequency precision and the coherence of a manipulation laser on typical experimental timescales are of the highest importance. Furthermore, future applications exploiting the quantum nature of optical links have recently been envisioned, allowing for the realization of a ``quantum network of clocks'' with quantum-enhanced capabilities \cite{ref38}. This upgrade to the ``quantum level'' would also represent a major breakthrough from a non-metrological point of view, allowing for quantum optics experiments over truly macroscopic distances, for both foundational tests and innovative technological applications.

\subsection*{Acknowledgements}

\noindent
We thank P. C. Pastor, P. De Natale, P. Lombardi and C. Sias for discussions and experimental contributions and A. Godone and M. Pizzocaro for a careful reading of the manuscript. This work has been supported by research grants MIUR PRIN2012 project AQUASIM, EU FP7 project SIQS and MIUR "Progetti Premiali" project LIFT.

\subsection*{Author contributions}

\noindent
(*) These authors contributed equally to the work.\newline
C. C., M. F, A. M., G. A. C., F. L., D. C. realized and operated the link and Cs fountain primary frequency standard and performed the measurements at INRIM.\newline
G. C., L. L., F. P., M. S. d. C., M. M., G. P., L. F., J. C., M. I. set up the regeneration stage, took care of experimental spectroscopic apparatus and performed the measurements at LENS.\newline
All authors contributed to the analysis of the data, to the discussion of the results, and to the writing of the manuscript.

% \section*{Author Information}
% The authors declare no competing financial interests. Correspondence and requests for materials should be addressed to L.F. 

\renewcommand{\thefigure}{S\arabic{figure}}
 \setcounter{figure}{0}
\renewcommand{\theequation}{S.\arabic{equation}}
 \setcounter{equation}{0}
 \renewcommand{\thesection}{S.\Roman{section}}
\setcounter{section}{0}
\renewcommand{\thetable}{S\arabic{table}}
 \setcounter{table}{0}

\onecolumngrid

\newpage

%%%%%%%%%%%%%%%%% METHODS %%%%%%%%%%%%%%%%%%%%%%%%%

% \begin{center}
% {\bf \large Supplemental Material for\\
% ``XXXXX''}\\
% \bigskip
% Authors
% \end{center}
% 
% \bigskip
\twocolumngrid

\subsection*{\large Methods}
\vspace{3mm}

\noindent {\bf Dissemination of an absolute optical reference.}
The ultrastable laser light that is sent from INRIM to LENS comes from a commercial 1542-nm fiber laser, frequency-stabilized to a high-finesse Fabry-Perot cavity using the Pound-Drever-Hall technique. Its relative frequency instability is $5 \times 10^{-15}$ at 1 s in terms of Allan deviation and it has a residual drift of $5 \times 10^{-15}$/s. To allow its use as an absolute frequency reference, it is phase-locked to a hydrogen maser using a fiber frequency comb. The beatnote between the 1542-nm laser and the comb is recorded with a commercial phasemeter at a sampling rate of 10 Hz and stabilized by applying a discrete frequency correction on an acousto-optic modulator (AOM) every 100 ms, on a bandwidth of 40 mHz. The phase-lock to the hydrogen maser slightly degrades the short-term stability, leading to an Allan deviation of $1.5\times 10^{-14}$ at 1 s measurement time. The laser is then sent to LENS through a 642-km-long fiber, phase-stabilized according to the Doppler noise cancellation scheme. A complete description of the setup can be found in \cite{ref8}; the round-trip beatnote is continuously tracked by two independent voltage-controlled oscillators (VCO); one of them feeds the PLL used for the link stabilization, the other is used for the phase-slips detection. To this purpose, the VCOs outputs are synchronously counted and all measurements differing by more than 0.4 Hz identify the occurrence of phase jumps. This threshold is chosen according to the minimum slips amplitude, i.e. 1 cycle, that is detected as a 1-Hz frequency outlier on a gate time of 1 s. Thanks to the low phase-slips rate (few per hour) and to their low amplitude (typically, less than 2 cycles), a thorough removal of the phase-slips is not necessary in this experiment, and it is enough to keep trace of their occurrence and amplitude. An overall uncertainty is assigned off-line to the disseminated frequency by measuring the discrepancy between the two VCO-frequencies.
\vspace{3mm}

\noindent {\bf Regeneration and fiber phase-noise cancellation.}
At the output stage of the fiber link at LENS, a clean-up narrow-line diode laser, driven by a home-made low-noise current driver, is phase-locked to the incoming radiation on a bandwidth of 100 kHz through direct current feedback on the laser chip. This stage amplifies the incoming link signal ($\sim 40$ nW), rejecting the wideband noise due to the build-up of Amplified Spontaneous Emission of the 9 fiber amplifiers used in the link path. Part of this light is sent back through the link for fiber-noise cancellation (see previous section). The final S/N ratio after regeneration and filtering is more than 50 dB on a 10 MHz bandwidth and the available power exceeds 5 mW. This grants the possibility for different end-users to withdraw a portion of the light, in a ``point-to-star'' dissemination scheme, currently connecting 4 different end-users located in 200 m range from the regeneration node. In the present work, the regenerated light is transferred to the final end-user lab via an additional 150-m-long, non-stabilized fiber. Even assuming temperature drifts of the order of 1 K/hour (worst-case scenario), the excess noise introduced by this secondary link is affecting the stability only below the $10^{-15}$ level on our experimental timescales. 
\vspace{3mm}

\noindent {\bf Analog/digital frequency lock to remote reference.}
In the end-user lab, a frequency comb serves as a frequency bridge between the 1542-nm regenerated reference laser and the 1156-nm probe laser. The 777597th tooth of the comb spectrum is beaten with the incoming radiation. The beatnote is stabilized on a bandwidth of 250 kHz by adjusting the comb repetition rate with an intra-cavity electro-optic modulator. Also, a portion of the comb spectrum around 1156 nm is separately filtered, amplified and beaten against the local cavity-stabilized laser. This beatnote is acquired with a frequency counter over a 30 s gate time and stabilized by a digital feedback loop. The latter relies on a PC-based proportional-integral feedback scheme, and  controls a Direct Digital Synthesis board, so that the frequency of the frequency-doubled laser light at 578 nm can be adjusted through an AOM before illuminating the atomic sample (see Fig. 1).
\vspace{3mm}

\noindent {\bf Geodetic corrections.}
The orthometric height above the geoid in Turin at the caesium fountain location is 238.7(1) m and the gravity correction is limited by the value of the geoid at $1 \times 10^{-17}$ in terms of equivalent relative frequency. The orthometric height in Florence is 40.3(1) m; hence the frequency correction, using the sensitivity of $1.09\times 10^{-16}$/m in the near-Earth approximation, is $4.39(1) \times 10^{-15}$ or 2.277(5) Hz. Incidentally, if we consider that the two orthometric heights have been evaluated on the same geopotential model \cite{italgeo}, the uncertainty could be further reduced as the geoid potential is rejected, and we can directly consider the gravity potential difference of the two locations. In this case, the height difference uncertainty is reduced to just 3 cm, equivalent to $3 \times 10^{-18}$ relative frequency uncertainty or 2 mHz.


\begin{thebibliography}{99}

\bibitem{ref1}
Lombardi, M. The Use of GPS Disciplined Oscillators as Primary Frequency Standards for Calibration and Metrology Laboratories, {\it NCSLI Measure J. Meas. Sci.} {\bf 3} (3), 56--65 (2008).

\bibitem{ref3}
Ludlow, A. D., Boyd, M. M., Ye, J. \& Schmidt, P. O. Optical atomic clocks. {\it Rev. Mod. Phys.} {\bf 87}, 637--701 (2015).

\bibitem{ref4}
Ushijima, I., Takamoto, M., Das, M., Ohkubo, T. \& Katori H. Cryogenic optical lattice clocks. {\it Nature Photon.} {\bf 9}, 185--189 (2015). 

\bibitem{ref5}
Bloom, B. J., Nicholson, T. L., Williams, J. R., Campbell, S. L., Bishof, M., Zhang, X., Zhang, W., Bromley S. L. \& Ye, J. An optical lattice clock with accuracy and stability at the $10^{-18}$ level. {\it Nature} {\bf 506}, 71--75 (2014). 

\bibitem{ref6}
Hinkley, N., Sherman, J. A., Phillips, N. B., Schioppo, M., Lemke, N. D., Beloy, K., Pizzocaro, M., Oates, C. W. \& Ludlow, A. D. An atomic clock with $10^{-18}$ instability, {\it Science} {\bf 341}, 1215--1218 (2013).

\bibitem{epl2015}
Calonico, D., Inguscio, M. \& Levi, F. Light and the distribution of time. {\it EPL} {\bf 110}, 40001 (2015).

\bibitem{ref7}
Droste, S., Ozimek, F., Udem, T., Predehl, K., H{\"a}nsch, T. W., Schnatz, H., Grosche, G. \& Holzwarth, R. Optical-frequency transfer over a single-span 1840 km fiber link. {\it Phys. Rev. Lett.} {\bf 111}, 110801 (2013).


\bibitem{ref8}
Calonico, D., Bertacco, E. K., Calosso, C. E., Clivati, C., Costanzo, G. A., Frittelli, M., Godone, A., Mura, A., Poli, N., Sutyrin, D. V., Tino, G. M., Zucco, M. E. \& Levi, F. High-accuracy coherent optical frequency transfer over a doubled 642-km fiber link. {\it Appl. Phys. B} {\bf 117}, 979-–986 (2014).

\bibitem{ref9}
Lopez, O., Haboucha, A., Chanteau, B., Chardonnet, C., Amy-Klein, A. \& Santarelli, G. Ultra-stable long distance optical frequency distribution using the Internet fiber network. {\it Opt. Express} {\bf 20}, 23518--23526 (2012).

\bibitem{ref10}
Fujieda, M., Kumagai, M., Nagano, S., Yamaguchi, A., Hachisu, H. \& Ido, T. All-optical link for direct comparison of distant optical clocks. {\it Opt. Express} {\bf 19}, 16498--16507 (2011).

\bibitem{ref11}
Yamaguchi, A., Fujieda, M., Kumagai, M., Hachisu, H., Nagano, S., Li, Y., Ido, T., Takano, T., Takamoto, M. \& Katori, H. Direct comparison of distant optical lattice clocks at the $10^{-16}$ uncertainty. {\it Appl. Phys. Express} {\bf 4}, 082203 (2011).

\bibitem{ref12}
Matveev, A., Parthey, C. G., Predehl, K., Alnis, J., Beyer, A., Holzwarth, R., Udem, T., Wilken, T., Kolachevsky, N., Abgrall, M., Rovera, D., Salomon, C., Laurent, P., Grosche, G., Terra, O., Legero, T., Schnatz, H., Weyers, S., Altschul, B. \& H{\"a}nsch, T. W. Precision measurement of the hydrogen 1S2S frequency via a 920-km fiber link. {\it Phys. Rev. Lett.} {\bf 110}, 230801 (2013).

\bibitem{ref13}
Clivati, C., Costanzo, G. A., Frittelli, M., Levi, F., Mura, A., Zucco, M., Ambrosini, R., Bortolotti, C., Perini, F., Roma M. \& Calonico, D. A coherent fiber link for very long baseline interferometry. {\it IEEE Trans. Ultrason. Ferroelectr. Freq. Control} {\bf 62}, 1907--1912 (2015).

\bibitem{ref14}
Droste, S., Grebing, C., Leute, J., Raupach, S. M. F., Matveev, A., H{\"a}nsch, T. W., Bauch, A., Holzwarth, R. \& Grosche, G. Characterization of a 450-km baseline GPS carrier-phase link using an optical fiber link. {\it New J. Phys.} {\bf 17}, 083044 (2015).

\bibitem{lisdat15}
Lisdat, C. \textit{et al.} A clock network for geodesy and fundamental science. {\it arXiv:1511.07735} (2015).

\bibitem{ref15}
Bjerhammar, A. On a relativistic geodesy. {\it B. Geod.} {\bf 59}, 207--220 (1985).

\bibitem{ref16}
Argence, B., Chanteau, B., Lopez, O., Nicolodi, D., Abgrall, M., Chardonnet, C., Daussy, C., Darqui\'e, B., Le Coq Y. \& Amy-Klein, A. Quantum cascade laser frequency stabilization at the sub-Hz level. {\it Nature Photon.} {\bf 9}, 456--460 (2015).

\bibitem{gorshkov2010}
Gorshkov, A. V., Hermele, M., Gurarie, V., Xu, C., Julienne, P. S., Ye, J., Zoller, P., Demler, E., Lukin, M. D. \& Rey, A. M. Two-orbital $SU(N)$ magnetism with ultracold alkaline-earth atoms. {\it Nature Phys.} {\bf 6}, 289--295 (2010).

\bibitem{cazalilla2014}
Cazalilla, M. A. \& Rey, A. M. Ultracold Fermi gases with emergent SU($N$) symmetry. {\it Rep. Prog. Phys.} {\bf 77}, 124401 (2014).

\bibitem{taie2012}
Taie, S., Yamazaki, R., Sugawa, S. \& Takahashi, Y. An SU(6) Mott insulator of an atomic Fermi gas realized by large-spin Pomeranchuk cooling.
{\it Nature Phys.} {\bf 8}, 825--830 (2012).

\bibitem{pagano2014}
Pagano, G., Mancini, M., Cappellini, G., Lombardi, P., Sch{\"a}fer, F., Hu, H., Liu, X.-J., Catani, J., Sias, C., Inguscio, M. \& Fallani, L. A one-dimensional liquid of fermions with tunable spin. {\it Nature Phys.} {\bf 10}, 198--201 (2014). 

\bibitem{mancini2015}
Mancini, M., Pagano, G., Cappellini, G., Livi, L., Rider, M., Catani, J., Sias, C., Zoller, P., Inguscio, M., Dalmonte, M. \& Fallani, L. Observation of chiral edge states with neutral fermions in synthetic Hall ribbons. {\it Science} {\bf 349}, 1510--1513 (2015).

\bibitem{daley2008}
Daley, A. J., Boyd, M. M., Ye, J. \& Zoller, P. Quantum computing with alkaline earth atoms. {\it Phys. Rev. Lett.} {\bf 101}, 170504 (2008).

\bibitem{daley2011}
Daley, A. J. Quantum computing and quantum simulation with Group-II atoms. {\it Quantum Inf. Process.} {\bf 10}, 865--884 (2011).

\bibitem{cappellini2014}
Cappellini, G., Mancini, M., Pagano, G., Lombardi, P., Livi, L., Siciliani de Cumis, M., Cancio, P., Pizzocaro, M., Calonico, D., Levi, F., Sias, C., Catani, J., Inguscio, M. \& Fallani, L. Direct Observation of Coherent Interorbital Spin-Exchange Dynamics. {\it Phys. Rev. Lett.} {\bf 113}, 120402 (2014).

\bibitem{scazza2014}
Scazza, F., Hofrichter, C., H{\"o}fer, M., De Groot, P. C., Bloch I. \& F{\"o}lling, S. Observation of two-orbital spin-exchange interactions with ultracold SU(N)-symmetric fermions. {\it Nature Phys.} {\bf 10}, 779--784 (2014).

\bibitem{zhang2014}
Zhang, X., Bishof, M., Bromley, S. L., Kraus, C. V., Safronova, M. S., Zoller, P., Rey, A. M. \& Ye, J. Spectroscopic observation of SU($N$)-symmetric interactions in Sr orbital magnetism. Science {\bf 345}, 1467--1473 (2014).

\bibitem{pagano2015}
Pagano, G., Mancini, M., Cappellini, G., Livi, L., Sias, C., Catani, J., Inguscio, M. \& Fallani, L. A strongly interacting gas of two-electron fermions at an orbital Feshbach resonance. Preprint arXiv:1509.04256 (2015).

\bibitem{hofer2015}
H{\"o}fer, N., Riegger, L., Scazza, F., Hofrichter, C., Fernandes, D. R., Parish, M. M., Levinsen, J., Bloch, I. \&  F{\"o}lling, S. Observation of an orbital interaction-induced Feshbach resonance in 173-Yb. Preprint arXiv:1509.04257 (2015).

\bibitem{ref28}
Hoyt, C. W., Barber, Z. W., Oates, C. W., Fortier, T. M., Diddams, S. A. \& Hollberg, L. Observation and absolute frequency measurements of the $^1$S$_0$---$^3$P$_0$ optical clock transition in neutral ytterbium. {\it Phys. Rev. Lett.} {\bf 95}, 083003 (2005).

\bibitem{ref31}
Clivati, C., Mura, A., Calonico, D., Levi, F., Costanzo, G. A., Calosso, C. E. \& Godone, A. Planar-waveguide external cavity laser stabilization for an optical link with $10^{-19}$ frequency stability. {\it IEEE Trans. Ultrason. Ferroelectr. Freq. Control} {\bf 58}, 2582--2587 (2011).

\bibitem{ref32}
Levi, F., Calonico, D., Calosso, C. E., Godone, A., Micalizio, S. \& Costanzo, G. A. Accuracy evaluation of ITCsF2: a nitrogen cooled caesium fountain. {\it Metrologia} {\bf 51}, 270--284 (2014). 

\bibitem{ref29}
Pizzocaro, M., Costanzo, G. A., Godone, A., Levi, F., Mura, A., Zoppi, M. \& Calonico, D. Realization of an ultrastable 578-nm laser for an Yb lattice clock. {\it IEEE Trans. Ultrason. Ferroelectr. Freq. Control} {\bf 59}, 426--431 (2012).

\bibitem{ref30}
Cappellini, G., Lombardi, P., Mancini, M., Pagano, G., Pizzocaro, M., Fallani, L. \& Catani, J. A compact ultranarrow high-power laser system for experiments with 578 nm ytterbium clock transition. {\it Rev. Sci. Instrum.} {\bf 86}, 073111 (2015).

\bibitem{ref33}
Williams, P. A., Swann, W. C. \& Newbury, N. R. High-stability transfer of an optical frequency over long fiber-optic links. {\it J. Opt. Soc. Am. B} {\bf 25}, 1284--1293 (2008).

\bibitem{derevianko}
Porsev, S. G., Derevianko, A. \& Fortson, E. N. Possibility of an optical clock using the 6$^1$S$_0$ $\rightarrow$ 6$^3$P$_0^0$ transition in $^{171,173}$Yb atoms held in an optical lattice. {\it Phys. Rev. A} {\bf 69}, 021403 (2004).

%\bibitem{Lemke09}
%Lemke, N. D., Ludlow, A. D., Barber, Z. W., Fortier, T. M., Diddams, S. A., Jiang, Y., Jefferts, S. R., Heavner, T. P., %Parker, T. E. \& Oates, C. W. Spin-1/2 optical lattice clock. {\it Phys. Rev. Lett.} {\bf 103}, 063001 (2009).

\bibitem{ref35}
Sherman, J. A., Lemke, N. D., Hinkley, N., Pizzocaro, M., Fox, R. W., Ludlow, A. D. \& Oates, C. W. High-accuracy measurement of atomic polarizability in an optical lattice clock. {\it Phys. Rev. Lett.} {\bf 108}, 153002 (2012).

\bibitem{poli08}
Poli, N., Barber, Z. W., Lemke, N. D., Oates, C. W., Ma, L. S., Stalnaker, J. E., Fortier, T. M., Diddams, S. A., 
Hollberg, L., Bergquist, J. C., Brusch, A., Jefferts, S., Heavner, T \& Parker, T. Frequency evaluation of the doubly forbidden $^1$S$_0$---$^3$P$_0$ transition in bosonic $^{174}$Yb {\it Phys. Rev. A} {\bf 77}, 050501 (2008).

\bibitem{ref36}
Calonico, D., Cina, A., Bendea, I. H., Levi, F., Lorini, L. \& Godone, A. Gravitational redshift at INRIM. {\it Metrologia} {\bf 44}, L44--L48 (2007).

\bibitem{ref37}
Cina, A. Private communication (2015).

\bibitem{zhang2015}
Zhang, R., Cheng, Y., Zhai, H. \& Zhang, P. Orbital Feshbach Resonance in Alkali-Earth Atoms. {\it Phys. Rev. Lett.} {\bf 115}, 135301 (2015).

\bibitem{ref38}
K{\'o}m{\'a}r, P., Kessler, E. M., Bishof, M., Jiang, L., S{\o{}}rensen, A. S., Ye, J. \& Lukin, M. D. A quantum network of clocks. {\it Nature Phys.} {\bf 10}, 582--587 (2014).


\bibitem{italgeo}
Barzaghi, R., Borghi, A., Carrion, D. \& Sona, G. Refining the estimate of the Italian quasi-geoid. {\it Bollettino di Geodesia e Scienze Affini} {\bf 66}, 145--149 (2007).

\end{thebibliography}
\end{document}